\documentstyle[aps,epsfig,multicol,fancyheadings]{revtex}
\begin{document}
%%%%%%%%%%%%%%%%%%%%%%%%%%%%%%%%%%%%%%%%%%%%%%%%%%%%%%%%

\title{Scaling of interfaces in brittle fracture and perfect
plasticity}

\author{Eira~T.\ Sepp\"al\"a, Vilho~I.\ R\"ais\"anen, 
and Mikko~J.\ Alava}

\address{Helsinki University of Technology, Laboratory of
Physics,\\ P.O.Box 1100, FIN-02015 HUT, Finland}
\date{\today}

\maketitle

\begin{abstract}

The roughness properties of two-dimensional fracture surfaces as
created by the slow failure of random fuse networks are considered and
compared to yield surfaces of perfect plasticity with similar
disorder. By studying systems up to a linear size $L=350$ it is found
that in the cases studied
the fracture surfaces exhibit self-affine scaling with a
roughness exponent close to 2/3, which is asymptotically exactly true
for plasticity though finite-size effects are evident for both. The
overlap of yield or minimum energy and fracture surfaces with exactly
the same disorder configuration is shown to be a decreasing function
of the system size and to be of a rather large magnitude for all cases
studied. The typical ``overlap cluster'' length between pairs of such
interfaces converges to a constant with $L$ increasing.

\end{abstract}

\noindent {\it PACS \# \ \ 62.20.Mk, 62.20.Fe, 05.40.-a, 81.40Np}

\begin{multicols}{2}[]
\narrowtext

\section{Introduction}\label{intro}

Roughness of fracture surfaces (FS) is a currently topical problem
that has opened up surprising connections between engineering and
weakly understood questions of statistical physics. The simple
question why and how a crack surface becomes rough dodges easy answers
since there is a multitude of experimental facts and ways for cracks
to develop or propagate. One simplification, adopted in this work is
to neglect cases in which the prevalent feature is the propagation of
``fast'' cracks in the favor of slow, adiabatic crack formation.  The
questions we address here are related to how disorder affects crack
surfaces and how interfaces born out of different load-elongation
responses are related. Disorder is present in materials at all length
scales in the form of atomic impurities, dislocations, grain
boundaries and so forth.

No generally accepted picture exists yet of how slow cracks are formed
and how it relates to crack interfaces \cite{Bourev}. In three
dimensions there are indications that the cracks become above a certain
intermediate length scale self-affine so that the roughness exponent
$\zeta$ is close to 0.8. Moreover the physics of crack advancement
indicates that the generic features of phase transitions of driven
lines (crack fronts in three dimensions) become relevant
\cite{bouprl}. Quantitative agreement is however missing. For slow
fracture in two dimensions (2D) and at small length scales in 3D the
interface scaling may be different in that the exponents are close to
those of the minimum energy (ME) interface. These are the same as for
the Random Exchange Ising Model (REIM) domain walls at zero
temperature, and have therefore the exact value $\zeta=2/3$ in
2D~\cite{HaH95} and the approximate value $0.41\pm0.01$ in
3D~\cite{Fis86,Middle,Mikko}. The physics involved is simple: the
crack minimizes up to the pertinent length scale the surface energy
$E$ given by
\begin{equation}
E = \int d^{d-1}{\bf x} \, \left[ \Gamma \{ \nabla z({\bf x}) \}^2 
+ V_r\{{\bf x}, z({\bf x})\} \right],
\label{eqE}
\end{equation}
where the integral counts in two contributions. One arises from a
surface stiffness (proportional to $\Gamma$) due to the deviations
$\nabla z({\bf x})$ from a straight crack, and a second from a random
disorder potential with a two-point correlator $\langle V_r({\bf
x}',z') V_r({\bf x},z) \rangle$ where the disorder average is implied
and $({\bf x}',z')$,$({\bf x},z)$ denote two locations inside the
medium. The exponents quoted above are true in case the disorder has
point-like correlations. The fluctuations of the potential would in an
experiment correspond to a varying failure threshold or elastic
modulus etc., depending on the circumstances. The closeness of the
numerical values of the roughness exponents gives rise to the
intriguing question as to why should a slow fracture resemble a global
optimization like ground state domain walls?  The connection is
suggested by the fact that in arbitrary dimensions, lattice models
that describe scalar perfect plasticity can be exactly mapped to the
REIM domain wall problems. For brittle fracture or vectorial failure
problems in general the correspondence is not obvious.
Two-dimensional failure is special in that there is some experimental
evidence of the crack roughness scaling with the domain wall in REIM,
i.e. the so called directed polymer (DP), global roughness
exponent~\cite{Kertesz}.  This connection between global optimization
and fracture surfaces has also been made in 2D simulations of brittle
failure~\cite{Hansen,letteri}. In 3D, it is still unclear whether even
just numerical models show such a
universality~\cite{letteri,pitka,BaHa}.

In this paper we investigate in two dimensions the scaling properties
of slow fracture surfaces and compare them to minimum energy surfaces
with similar {\it a priori} disorder. We perform numerical simulations
of the random fuse network (RFN) model, which has been studied
extensively as a model of brittle failure of disordered materials
\cite{review,Dux,Kahng}. As a warm up, we consider extensive system 
properties such as fracture stress, fracture strain, and damage. For
the main case studied here, the dilution-type disorder, these are
found to be in good agreement with the critical defect -type arguments
proposed by Duxbury and co-workers~\cite{Dux} that imply logarithmic
scaling with system size.  Note that when comparing with ``reality''
this kind of models contain two assumptions: first, the stress
relaxation is supposed to be much faster than the stress rate
(an adiabatic failure). Second, one assumes that the energy released
by local crack formation is dissipated with no effect on the crack
propagation.

The paper starts with a short description of the numerical methods
used and the dynamics of the adiabatic crack formation in
Section~\ref{num}.  Section~\ref{scale} discusses the strength
properties of random fuse networks as a function of system size $L$.
There are a number of ways to characterize a posteriori a self-affine
interface. This is the main theme of Section~\ref{sample}, the topic
of showing that 2D brittle fracture interfaces have the DP-type
scaling. We demonstrate how both the so-called local width and the
statistical properties of ensembles of interfaces indicate a similar
kind of self-affine scaling. The scaling exponent is seen to be close
to the DP one, $\zeta=2/3$. The section also contains numerical data
for varying disorder strength, and in particular compares perfect
plasticity and brittle fracture by measuring the overlap of the
associated interfaces starting with the same disorder
configurations. This would be particularly relevant should it be so
that the fracture and yield interfaces use the same ``valley'' in the
landscape of the energies or thresholds.  The paper is finished with a
discussion in Section~\ref{discussion}.

\section{Creating the interfaces}\label{num}

\subsection{Numerical models}\label{model}
\begin{figure}[f]
\centerline{\epsfig{file=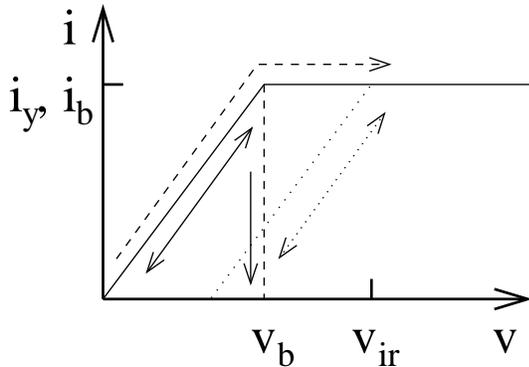,width=7cm}}
\caption{The voltage-current diagram of a fuse or a media. The
$\longleftrightarrow$ arrow describes an ideal elastic behavior. The
dashed line describes an ideal elastic break down at the critical
current $i_b$ with the corresponding voltage $v_b$. The dashed arrow
describes an perfect plasticity with yield current $i_y$, and the
dotted arrow describes an elastic-plastic behavior with an
irreversible yield strain $v_{ir}-v_b$ at $v_{ir}$. The current as a
function of the voltage with the corresponding elastic or plastic
behavior of the media may only increase or decrease in the directions
noted by the arrows.}
\label{fig1}
\end{figure}

Random fuse networks are electrical analogues of elasticity and
failure with disorder incorporated. One usually sets to mimic a
tensile test implying that the extensive thermodynamic parameters
become $V_{ext}$ and $I_{ext}$, external voltage and current,
respectively. These correspond to displacement and force in a real
experiment. To study brittle failure one defines the elements that
connect two nodes on an original lattice as {\it fuses}. These have a
linear voltage-current relationship until a breakdown current $i_b$,
see Fig.~\ref{fig1}.  A second choice would correspond to perfect
plasticity if one made the fuses such that the local current becomes
irreversibly constant at $i_y$ and stays so unless the local voltage
is reduced, in which case the conductivity becomes the original one
and there is a permanent {\it yield strain}.

In the following we use two different numerical techniques to study
both brittle and perfectly plastic RFN's. Brittle failure is studied
with standard adiabatic fracture iterations.  These consist of solving
the current balance in the system from Kirchoff's and Ohm's laws and
breaking after each iteration the most strained fuse (the criterion is
$\min(i_j / J_{c,j})$, where $i_j$ is the local current in each of the
fuses and $J_{c,j}$ is the local threshold). The currents and voltages
are found by solving the linear system of currents by the
conjugate-gradient method.

For perfect plasticity we use a mapping to minimum energy interfaces,
i.e. random exchange Ising domain walls in their ground state, where
exchange constants $J_{ij}$ between nearest neighboring spins are
random but non-negative.  In some cases we have exactly the same
quenched disorder (equal thresholds for failure $i_b$ and yielding
$i_y \equiv 2J_{ij}$ for each fuse) as for brittle failure, and in the
following the threshold for a fuse in both cases is denoted by
$J_c$. The simulations are done using combinatorial optimization:
finding the yield path (in 2D) is equivalent to the minimum
cut-maximum flow problem of network flows~\cite{Alavaetal} that
minimizes $\sum_{interface} i_y$.  This technique is more convenient
than the transfer matrix methods in that there are no restrictions
for the shape of the optimal path as overhangs and arbitrary
transverse steps are included in a natural fashion.

The typical choice for introducing disorder to a RFN is to pick the
failure currents $J_c$ from a prescribed probability distribution
$P(J_c )$.  The important issue is the behavior of $P$ for $J_c \simeq
0$ and for $J_c \rightarrow \infty$; the tails of the distribution are
known to have strong effects on the strength properties and damage
accumulation in the case of brittle fracture. For perfect plasticity
or directed polymers the case is much simpler in that for one
dimensional interfaces in $(1+1)$ dimensional systems such point-like
disorder is in the renormalization group sense always relevant. Thus
one expects always the same scaling properties in terms of interface
roughness and sample-to-sample interface energy fluctuations, these
correspond to yield stress fluctuations in plasticity. The amplitudes
are however non-universal and thus will depend on the exact form of
$P$.

In the following we study as typical examples the cases where $P(J_c)$
is a flat distribution ($P(J_c)= 1/(2\, \delta J)$ for $J_0 - \delta
J \leq J_c \leq J_0 + \delta J$) and where $P$ corresponds to
``dilution disorder''.  That is $P(J_c) = p \, \delta(J_c-1) +
(1-p) \, \delta(J_c)$.  The fraction of fuses that remain for
infinitesimal currents with dilution is denoted with $p$, which has a
value $p=0.8$ unless otherwise mentioned, like for the uniform
distribution case $\delta J/J_0 =1$.  The systems are chosen so that
the direction of macroscopic current flow is aligned in the $\langle
10 \rangle$ orientation of the square lattice having periodic
boundaries in the perpendicular direction. The systems are isomorphic,
i.e. $L_x = L_z$, and their sizes range from $L^2 =10^2$ to $350^2$
for brittle failure and to $1000^2$ for perfect plasticity,
respectively. The mean positions of the surfaces are not fixed, hence
they may sit anywhere in the system. The interfaces are defined in the
usual way so that in the case of overhangs, the so called
solid-on-solid approximation is used, i.e. the interface is found by
taking the furthest value of the interface with respect to a fixed end
of the network.  The number of realizations $N$ over which the
disorder-averaging is performed is limited by the CPU-time for the
simulations of the brittle fracture.  In the case of plasticity, the
technique used leads to a roughly linear scaling with respect the
number of fuses in a system regardless of the threshold distribution.
The number of different random realizations is shown in
Table~\ref{tabN} for the cases in which exactly the same random
networks are studied for both the brittle failure and the perfect
plasticity. If only the ME-surfaces are studied $N=200-5000$.

\subsection{Formation of the interfaces}
One should note that if fracture surfaces have non-trivial geometric
scaling properties and in particular resemble directed polymers (in
2D) this opens up several further questions: whether the outcome is
independent of disorder strength, whether the disorder is always
relevant for 2D minimum energy surfaces, and how the surface roughness
relates to other typical quantities.  The standard way of iterating
fracture in fuse networks, whether perfectly plastic or brittle, is
based on extremal dynamics.  The condition $\min(i_j / J_{c,j})$ for
the failure of the next element contains two effects: the disorder
through the threshold and the local current that depends on the
environment of the fuse.

For perfect plasticity, the information necessary for finding the
final yield surface is contained in the initial field $J_{c,j}$ due to
the monotonicity property noticed by Roux and
Hansen~\cite{HansenRoux}.  Even if one simulates the development of
the system as a serie of fuse network problems (the tangent problem),
the local current never decreases in a yielding process. Thus the
final yield surface equals a blocking configuration that can be
calculated from the original thresholds. This is related to the fact
that the surface is much faster to compute than the whole process by
considering it as an optimization problem for the interface: the
history of the whole process or system involves much more information.

For brittle fracture the monotonicity property is not true and thus no
direct mapping exists between the initial disorder and a quantity to
be minimized.  The mapping of perfect plasticity to fuse networks
makes it on the other hand clear that the difference between the
processes is smaller than it would seem at the first glance. This is
because in the tangent algorithm one has to solve a serie of adiabatic
failure problems with the local yield thresholds $i_y$ renormalized by
subtracting the current already passing through the fuse. Nonetheless
each failure iteration is affected by stress-enhancement effects
exactly as in a failure problem with the same fuses still intact. For
brittle failure, the implication of the stress-enhancements during the
failure process is that in order to obtain a minimum energy surface
(as defined by Eq.~(\ref{eqE})) the original disorder $i_{b,j}$ has to
be {\it renormalized}. That is, the thresholds or missing fuses
contain frozen-in information about how the field of local stresses
will develop and normalize the local thresholds $i_{b,j}$ in the
failure criterion. Considered in this light, it is sensible that the
brittle fracture surfaces are ``blocking paths'', too. Yet the
question remains whether the interfaces are still in the same {\it
universality class}: if the correlations in the renormalized disorder
become different enough from point-like correlations the interface
scaling properties will change. E.g.  columnar correlations ($V_r({\bf
x}, z)$ constant along $x_i$ or $z$) would be relevant in this
respect.

\section{Scaling of fracture}\label{scale}

A fracture can be contrasted with perfect plasticity also
by looking at extensive thermodynamic quantities. The standard
ones to consider are the damage $n_b$, the number
of fuses broken in total, and the failure current $I_b$ and
voltage $V_b$ as computed from the maximum current of the $IV$-curve.
For ``truly'' brittle failure this definition of $V_b$ agrees with
with that defined as the end point of the $IV$-curve. In the
failure of brittle fuse networks there is considerable evidence
for the relevance of critical defect -type effects. That is,
the defect with the largest current enhancement will dictate
the scaling of the current and voltage. For yield surfaces
one would have $I_y = E \sim L$, $\Delta E \sim L^\theta$ where
$\theta = 2 \zeta -1$ and $\zeta$ is the roughness exponent.
Thus the critical strength quantities, without the renormalization
discussed above, are supposed to have different scaling behavior
in plasticity and fracture.

\begin{figure}[f]
\centerline{\epsfig{file=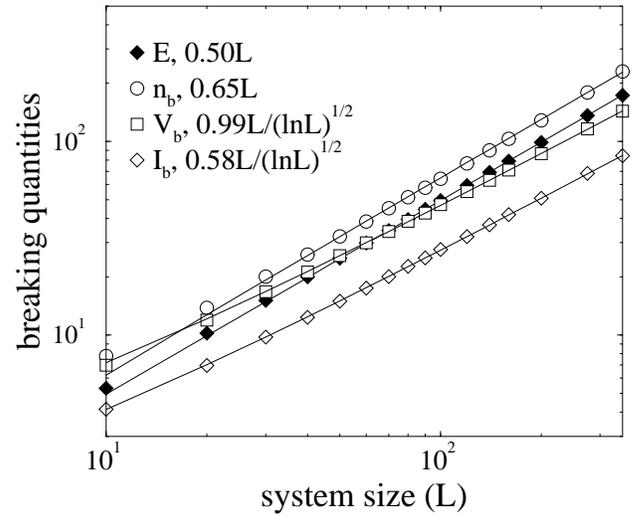,width=7cm,angle=-90}}
\caption{Scaling of energy of ME surfaces, closed diamonds, and fracture 
quantities: total damage, i.e. number of broken fuses $n_b$, open
circles; breaking voltage $V_b$ of the network, open squares; and
breaking current $I_b$ of the network, open diamonds, as a function of
the system size $L$. The disorder is dilution type with $p=0.8$.  The
number of realizations $N$ for the brittle failure case is shown in
Table~\ref{tabN}. For ME surfaces $N=5000$ for $L^2=10^2-50^2$,
$N=1000$ for $L^2=60^2-350^2$. The lines are least squares fits,
linear for $E$ and $n_b$; $L/\sqrt{\ln L}$ for $V_b$ and $I_b$, to the
data.}
\label{fig2}
\end{figure}

Fig.~\ref{fig2} shows the scaling of damage and the strength
quantities for dilution-type disorder, $p=0.8$. The lines in the
figure have been found with least-square-fits to data using the
assumptions of linear scaling for $n_b$ and for the other two
quantities the scaling $V_b,I_b\sim L/\sqrt{\ln L}$, which comes from
the extreme value statistics, i.e. Gumbel distribution, studied in the
fracture case by Duxbury and coauthors~\cite{Dux}. It is seen that the
scaling of the number of broken fuses is asymptotically very close to
a linear one. This means that the system still breaks in a brittle
mode for $p=0.8$.  For $V_b$ and $I_b$ the scaling in the whole regime
follows beautifully the $L/\sqrt{\ln L}$ scaling. Notice that the
surface energy of yield surfaces is in principle a lower limit for
$n_b$ and that both scale linearly. One sees that the energy of yield
surfaces or the lower limit for $n_b$ is lower by a constant factor
than $n_b$ of fracture surfaces. Similar behavior is visible in the
roughness values of the fracture and yield surfaces studied in the
next section.

\section{Scaling of interfaces}\label{sample}

\subsection{Global and local interface width}\label{local}

There are several ways to characterize the scaling properties of
interfaces. Consider the case where an interface is defined as a
function of ${\bf x}$ as $z({\bf x})$. The standard way of looking at
scaling properties is to calculate the interface width~\cite{Racz} or
standard deviation, i.e. the so-called root-mean-square (RMS)
roughness,
\begin{equation}
w = \left \langle \frac{1}{L^{d-1}} \sum^{L^{d-1}}_{{\bf x} = 1}
[z({\bf x}) - \bar{z}]^2 \right \rangle^{1/2},
\label{w_eq}
\end{equation}
where $\bar{z}$ denotes the mean position of an interface and $\langle
\, \rangle$ the disorder-average over the different random
configurations. If the interface is self-affine, $w$ should scale with
$L^\zeta$, $\zeta$ being the roughness scaling exponent. For
self-affine interfaces the scaling exponent $\zeta$ is expected to be
valid also for higher-order statistics. This is seen for the
height-height correlation functions \hbox{$G_k(l)= \langle |z({\bf x})
z({\bf x}+l)|^k \rangle $} and is applicable to the local width as
well. Note that there is no a priori reason to use a
Family-Vicsek-type of scaling Ansatz with a correlation
length~\cite{Barabasi}, since there is no dynamical length scale for
these interfaces. This is an assumption for brittle failure, and it
will be shown to hold by our data below, and is moreover exactly true
for perfect plasticity. It is also theoretically appealing since slow,
adiabatic failure does not involve any time scale.

The local width in two-dimensions is defined analogously to the global
interface roughness $w$ with
\begin{equation}
w_{loc}^2 (l) = \left \langle \frac{1}{l} \sum^l_{x = 1} [z(x) -
\bar{z}_l]^2 \right \rangle,
\label{w_loc}
\end{equation}
where the local interface height $\bar{z}_l$ is averaged over windows
of size $l \leq L$. One should note the obvious connection to $G_2$ that
exists both for the local and global definitions of interface width.

The advantage of using more complicated indicators of scaling is that
one can draw conclusions based on data in a much more limited system
size range than with the global interface width. Of course, such a
posteriori techniques are most commonly used in the context of
characterizing experimental fracture surfaces. Here we note the fact
that for small $L$ finite size effects make it rather difficult to
determine the roughness exponent (if one assumes the interfaces to be
truly self-affine to begin with). This is especially true for 3D
systems for which the computational costs become prohibitive easily.

\begin{figure}[f]
\centerline{\epsfig{file=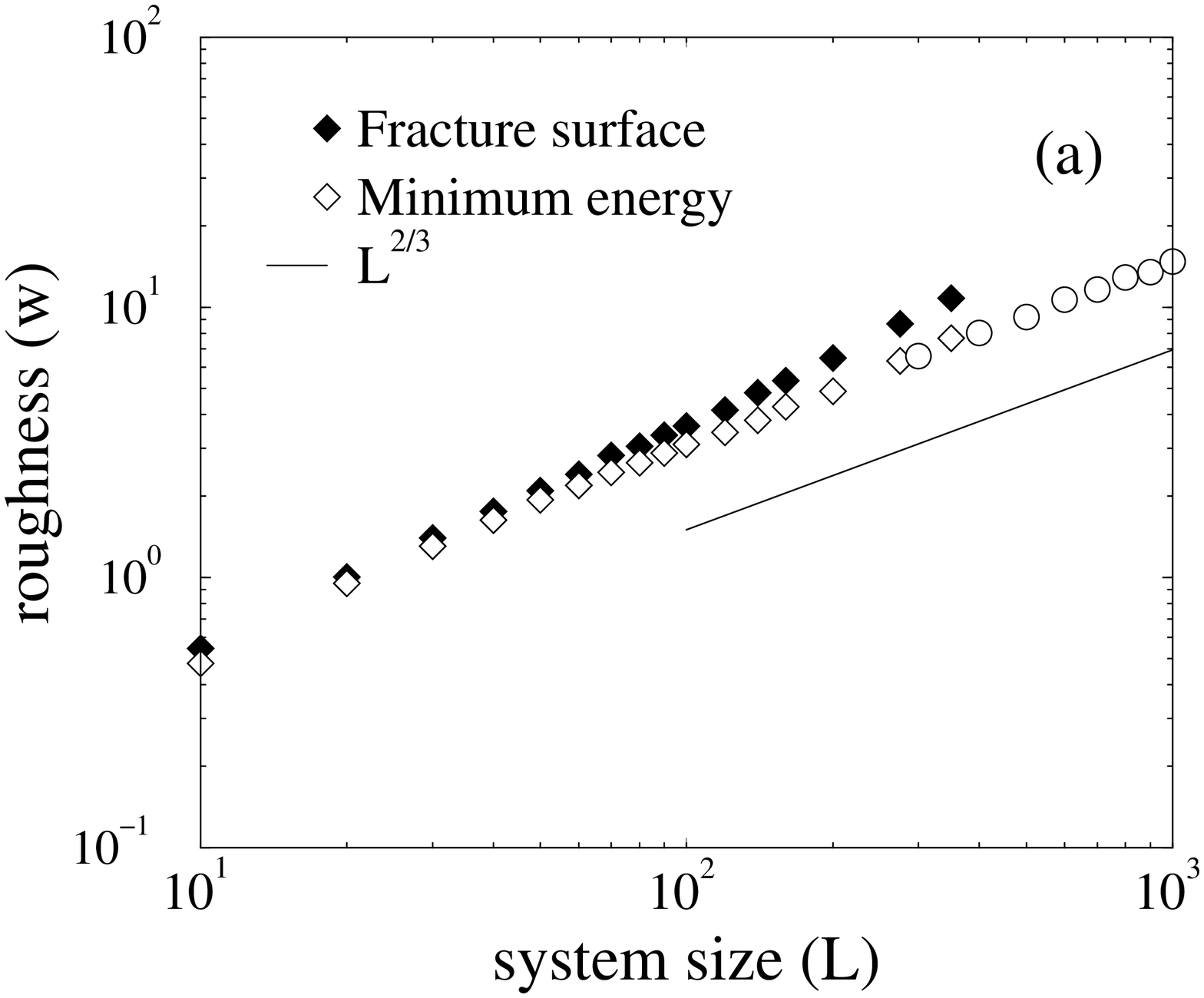,width=7cm,angle=-90}}
\centerline{\epsfig{file=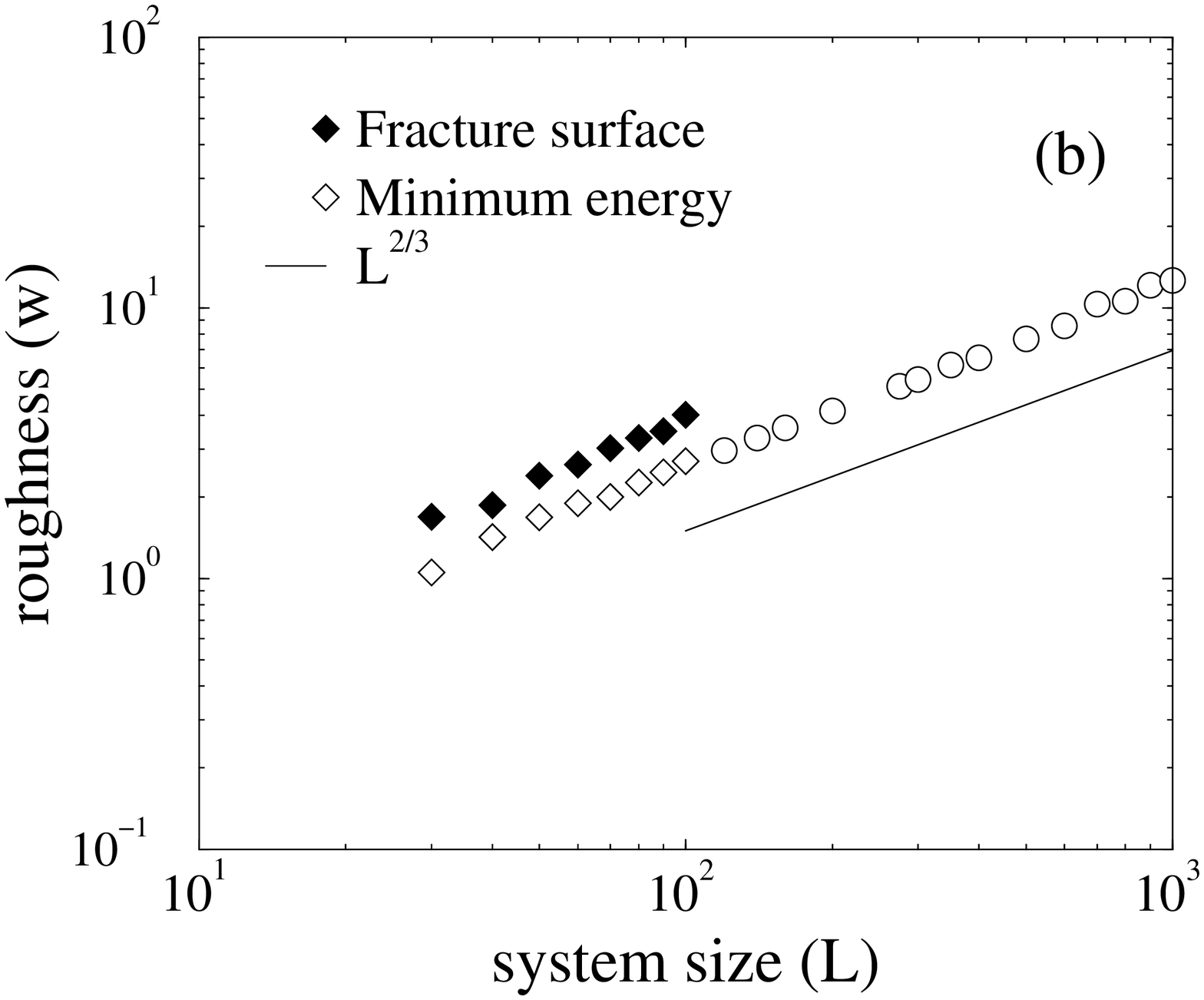,width=7cm,angle=-90}}
\caption{The interface width $w$ versus the system size $L$ for the
brittle fracture interfaces, closed diamonds, and the minimum energy
ones, open diamonds, from the same random networks. Open circles are
minimum energy interfaces from larger system sizes. The disorder is
(a) dilution-type with $p=0.8$ and (b) from uniform distribution of
fuse thresholds with $\delta J/J_0 =1$.  The number of different
random realizations for the exactly same fracture and yield surfaces
is shown in Table~\ref{tabN}. ME surfaces have $N$=1000 for
$L^2=300^2, 400^2$, $N$=500 for $L^2=500^2-700^2$, and $N$=200 for
$L^2=800^2-1000^2$ in (a), $N$=500 for $L^2=120^2-400^2$, and $N$=200
for $L^2=500^2-1000^2$ in (b). The lines are guides to the eye with a
slope $\zeta=2/3$.}
\label{fig3}
\end{figure}

The global interface roughness $w$ as a function of system size is
compared between directed polymers and brittle fracture interfaces
with the dilution-type disorder and uniform distribution of $J_c$ in
Fig.~\ref{fig3}.  As expected, for small $L$ the systems suffer from
finite size effects having exponent greater than $\zeta=2/3$ but
eventually the exponent becomes comparable to the value one obtains by
fitting a power-law to the large-$L$-data. Specifically, in the
dilution case, Fig.~\ref{fig3}(a), the effective exponent for the
fracture surfaces is $\zeta_{FS}\simeq 0.82$ and for the minimum
energy interfaces with exactly the same random threshold
configurations $\zeta_{ME,<}\simeq 0.74$.  For larger system sizes,
which we are able to study numerically only in the plasticity limit
with large enough number of realizations, $\zeta_{ME,>}\simeq
0.67$. For the fuses from the uniform random distribution,
Fig.~\ref{fig3}(b), the fracture surfaces scale with $\zeta_{FS}\simeq
0.73$ and the yield surfaces from the networks with exactly the same
random configurations have $\zeta_{ME,<}\simeq 0.74$. Hence the finite
sizes effects seem to be more similar between the processes than in
the dilution case. For the larger system sizes of minimum energy
interfaces $\zeta_{ME,>}\simeq 0.69$.

\begin{figure}[f]
\centerline{\epsfig{file=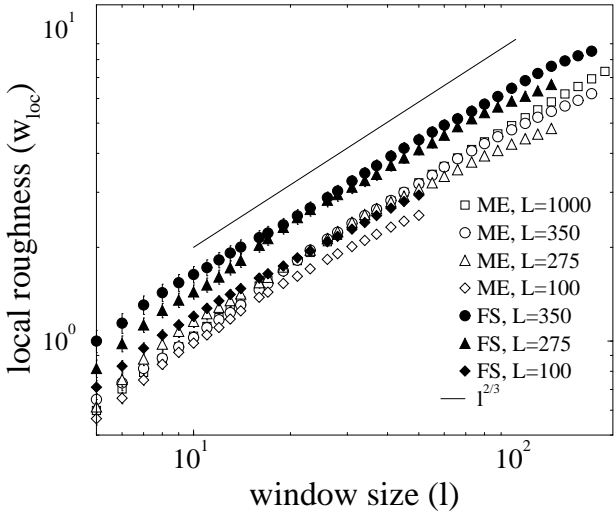,width=7cm,angle=-90}}
\caption{The local interface roughness $w_{loc}$ versus the window
size $l$ for the dilution type of disorder with $p=0.8$. The system
size $L^2=1000^2$ has $N=2000$, while for all the other system sizes
the data is from the same configurations as the data in
Fig.~\ref{fig3}(a). The line is a guide to the eye with a slope
$\zeta=2/3$. For the minimum energy interfaces the $\zeta=2/3$ scaling
is seen in a region $l<1/5L$ for larger system sizes, while
$L^2=100^2$ has a visible amplitude difference. The periodic
boundaries are used in the transverse direction of the external
voltage. The correlation between the local width of the elastic
fracture surfaces and plastic yield ones is clearly seen.}
\label{fig4}
\end{figure}

Fig.~\ref{fig4} compares the local width of directed polymers to the
brittle fracture interface for the dilution-type disorder with
$p=0.8$. For the directed polymers one sees that the $\zeta=2/3$
scaling is valid for larger system sizes in the region, where the
window size $l\leq 1/5 L$. With open boundary conditions the scaling
region would be larger. However, for smaller system size $L^2 = 100^2$
there is a visible amplitude difference compared to the larger system
sizes. The fracture surfaces show similar behavior but have larger
finite size effects when compared to the yield surfaces, and they have
a larger amplitude, too, than minimum energy surfaces in local
and in global width scaling.

Our result supports the conclusion of Ref.~\cite{letteri} that 2D
brittle fracture surfaces are in the directed polymer -universality
class ($\zeta = 2/3$), although due to the stronger finite size
effects the asymptotic region is harder to reach than for yield
surfaces.

\subsection{Roughness statistics}\label{distri}

\begin{figure}[f]
\centerline{\epsfig{file=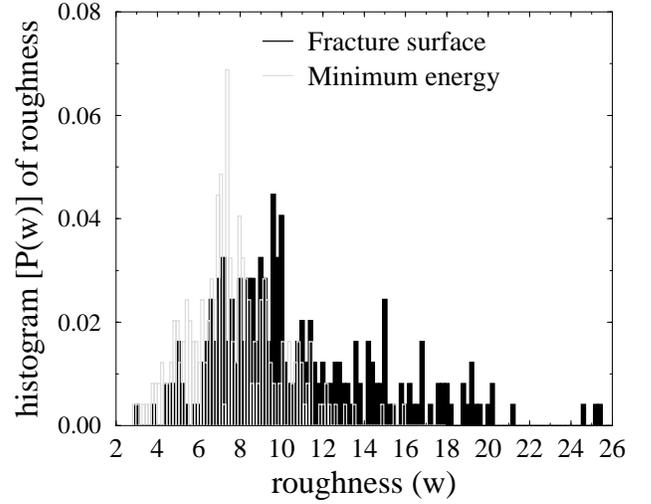,width=7cm,angle=-90}}
\caption{The histogram $P(w)$ of the roughness for the ideal
elastic and perfect plastic yield surfaces in systems of size
$L^2=350^2$ and dilution type of disorder with $p=0.8$.  The data is
from the same configurations as the data in Fig.~\ref{fig3}(a).}
\label{fig5}
\end{figure}

Next we address the higher-order statistics of fracture surfaces. For
directed polymers one knows that the end-point transverse deviation
distribution $z(x)$, ($=x(t)$ in the ordinary DP notation), is roughly
Gaussian for the standard case of one fixed, one free end (see
e.g.\cite{HaH95}) and follows a scaling form $P[z(L)] \sim f(z/L^\zeta
)$. One can likewise write down a scaling form for the interface
energy.  Next we assume that the brittle fracture interfaces obey
similar self-affine scaling and study the roughness probability
distribution $P(w,L)$ as a function of $L$ (we concentrate on the
$p=0.8$ case of the previous subsection).  Fig.~\ref{fig5} shows a
typical example of such distribution: it is not centered around zero
and is reminiscent of a log-normal or Poissonian distribution. This
can be understood qualitatively since the roughness $w$ is also a
measure of the non-zero maximum transverse displacement $z$.  The
figure includes a distribution for ME surfaces from the same systems,
too.

As one could see already in the previous subsection, the 2D fracture
surfaces are rougher than the minimum energy surfaces (i.e. assuming
self-affine scaling in both cases, the amplitude of the roughness
$w/L^\zeta$ is larger for FS). This is visible also here since the
distribution of $P(w)$ is not only wider but extends to higher values
for the fracture case. While assuming that $P(w)$ follows the same
self-affine scaling law as the $P[z(L)]$ we may study the
disorder-standard deviation
\begin{equation}
\sigma(w) = \left \langle (w - \bar{w})^2 \right \rangle^{1/2},
\label{w_std}
\end{equation}
where $w$ is from a single random system and $\bar{w}$ is the global
disorder-averaged roughness calculated using Eq.~(\ref{w_eq}).
$\sigma(w)$ scales with $L^{2/3}$, too, although the data is more
scattered due to the fact that higher-order statistics are always more
vulnerable to the finiteness of the statistics than the integral of
them.

\begin{figure}[f]
\centerline{\epsfig{file=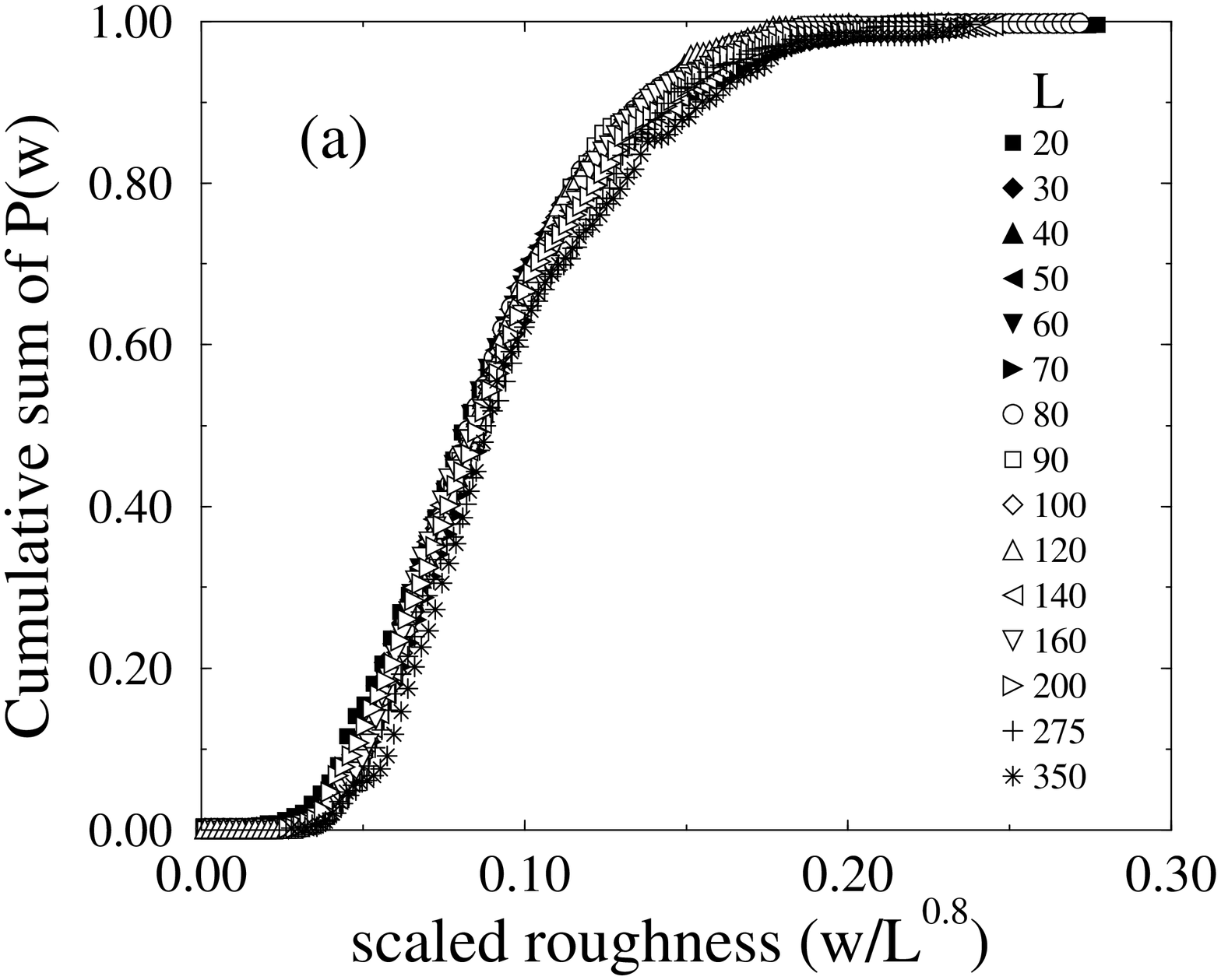,width=7cm,angle=-90}}
\centerline{\epsfig{file=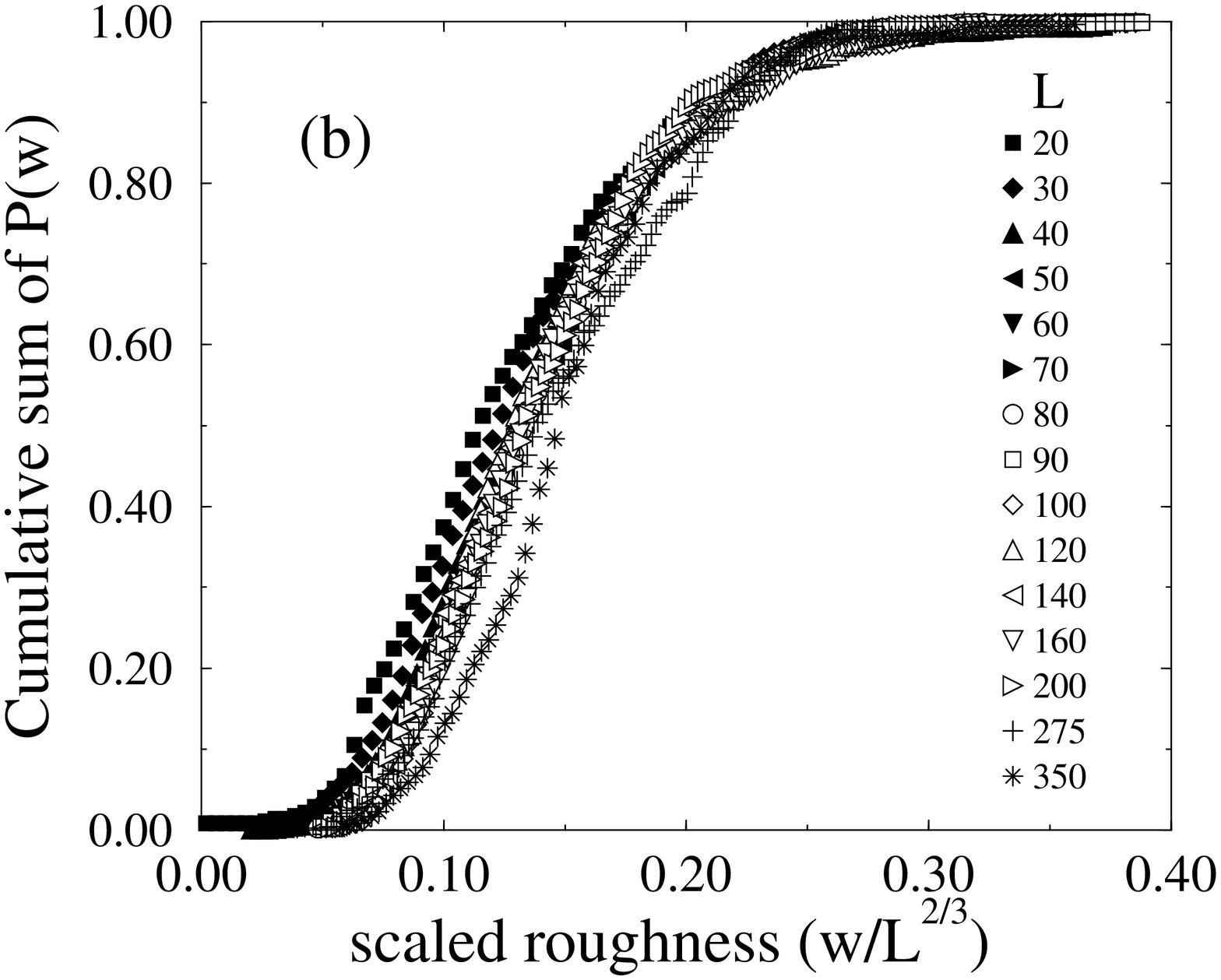,width=7cm,angle=-90}}
\caption{Cumulative sums of $P(w)$ for both fracture (a) and minimum energy
surfaces (b) for various system sizes with dilution type of disorder,
$p=0.8$.  The data is from the same configurations as the data in
Fig.~\ref{fig3}(a).  The data has been scaled with $L^{0.8}$ in (a)
and with $L^{2/3}$ in (b).}
\label{fig6}
\end{figure}

In Fig.~\ref{fig6} we collapse the data of the cumulative sums of the
distributions $P(w,L)$ for various $L$.  For both kinds of
interfaces the data-collapse is better with exponents $\zeta>2/3$,
which is due the finite size effects. For fracture surfaces $\zeta
= 0.8$, as in Fig.~\ref{fig3}(a), seems to work nicely, like for the
yield surfaces $\zeta = 0.74$ would be better than $\zeta = 2/3$.
In our opinion the figure justifies the assumption of an
asymptotically self-similar scaling of $P[w(L)] \sim f(w/L^\zeta)$.

\subsection{Scaling of overlap quantities}\label{overlap}

\begin{figure}[f]
\centerline{\epsfig{file=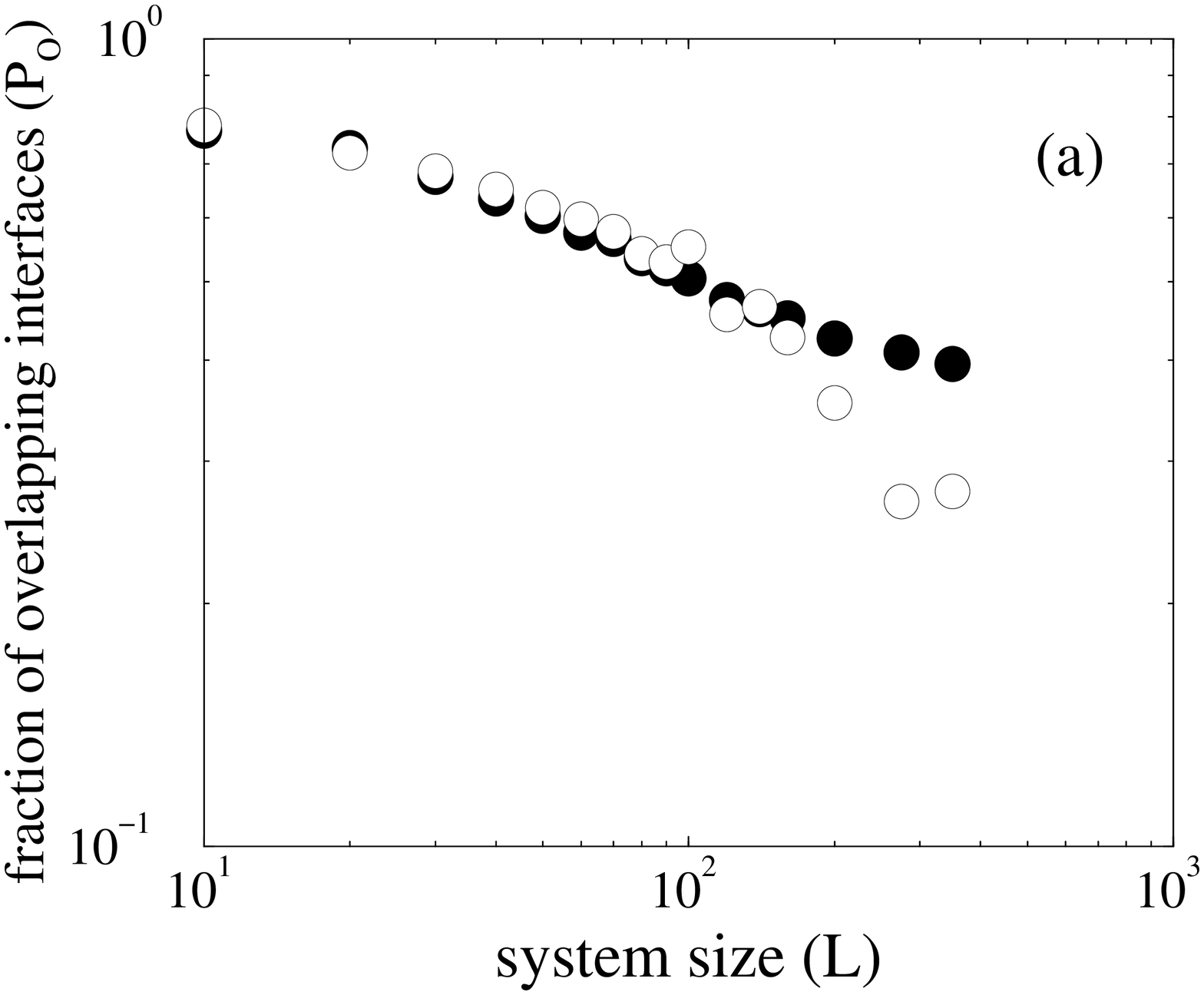,width=7cm,angle=-90}}
\centerline{\epsfig{file=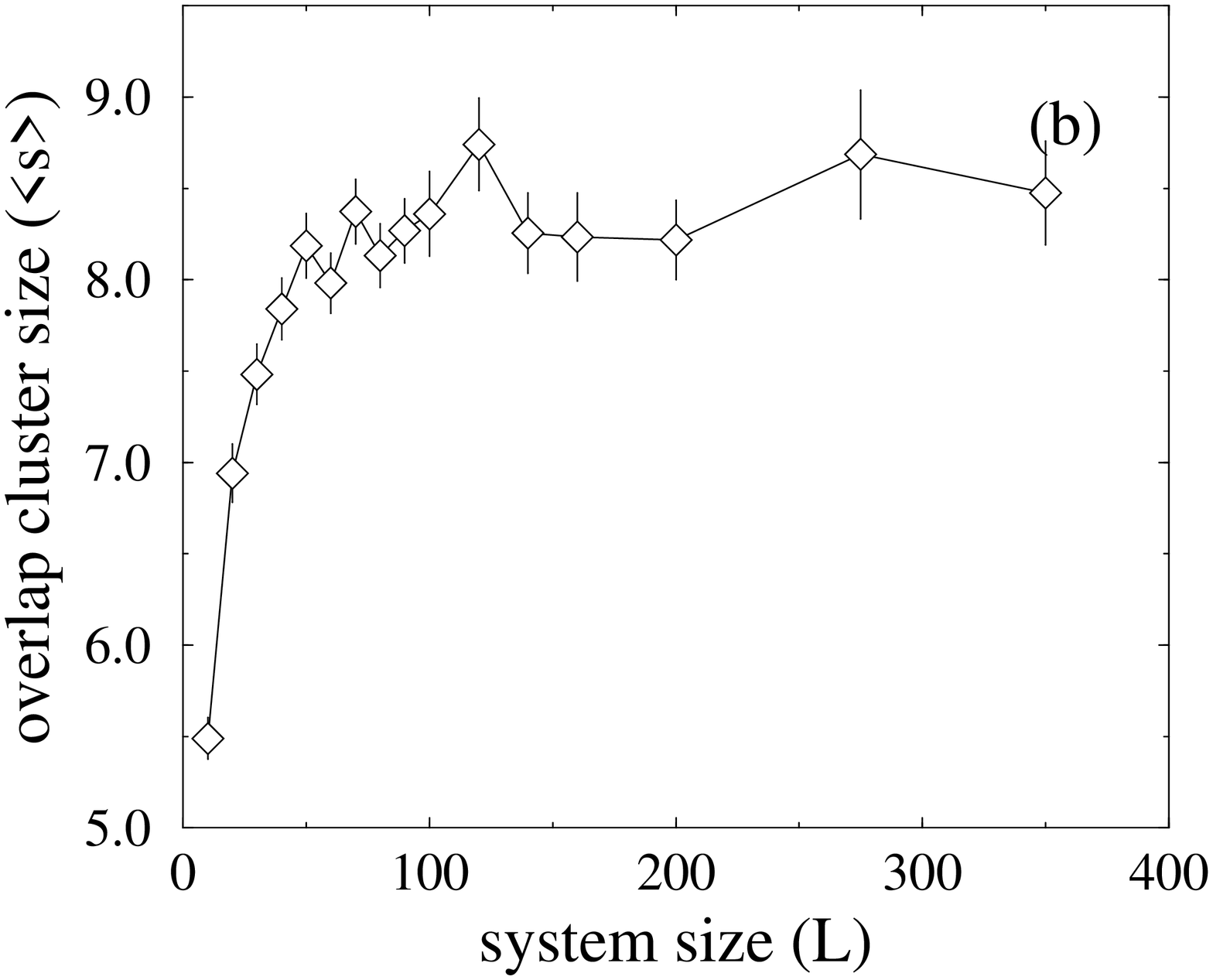,width=63mm,angle=-90}}
\centerline{\epsfig{file=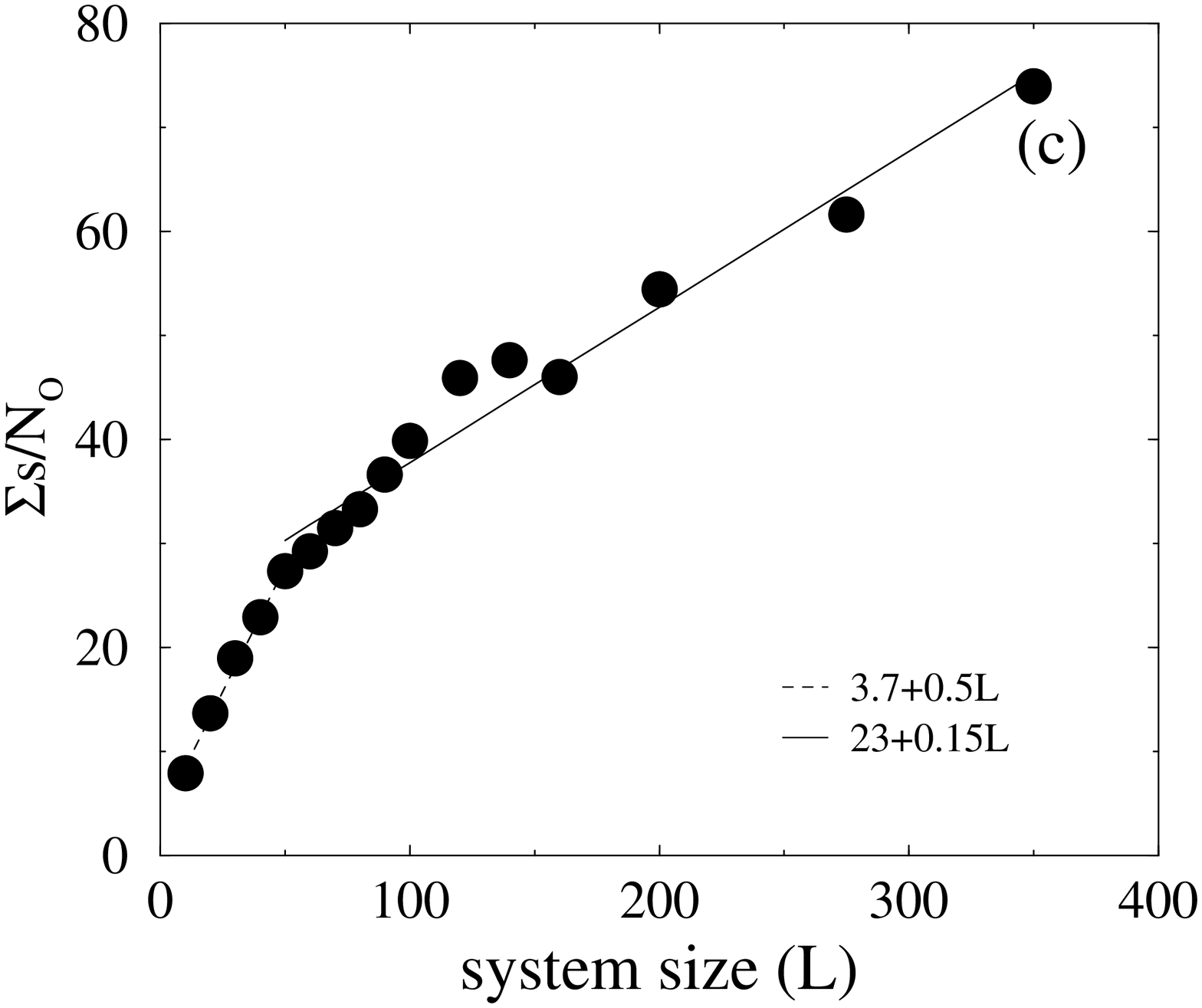,width=63mm,angle=-90}}
\caption{(a) Fraction of the disorder configurations, $P_O$, in which
the fracture surface and the minimum energy interface do have an
overlap, i.e. at least one common ($x,z$)-coordinate pair, open
circles. The disorder is dilution type with $p=0.8$.  The data is from
the same configurations as the data in Fig.~\ref{fig3}(a).  The closed
circles are for the comparison the value $P_{ran} =
7.5(w_{FS}+w_{ME})/L$ from the figure~\ref{fig3}(a), see the text for
details. (b) The average size $\langle s \rangle$ , i.e.  the number
of common neighboring ($x,z$)-coordinate pairs, of the overlapping
clusters as a function of the system size.  The overlap cluster size
saturates at $\langle s \rangle \simeq 8.5$.  (c) The total length of
overlap in configurations, which do have an overlap in their
interfaces, $\sum s/N_O$. The lines are linear least squares fits to
data.}
\label{fig7}
\end{figure}

The average overlap $P_O$ of fracture and minimum energy surfaces as a
function of the system size for the dilution case is seen in
Fig.~\ref{fig7}(a). Overlap is defined as the fraction of the disorder
realizations in which at least one $(x,z)$ coordinate pair is common
with the fracture and yield surfaces. Clearly the overlap is reduced
as a function of the system size. This is not surprising, because with
increasing system size the probability of the first breaks taking
place at the globally weakest place decreases. However, if one expects
the overlap quantity to originate from the result of depositing the
surfaces randomly (as particles of finite width on a 1D line segment
of length $L$), one obtains $P_{ran} = (A_{FS}w_{FS}+A_{ME}w_{ME})/L$.
$A_{FS}$ and $A_{ME}$ are prefactors needed to compute the typical
geometrical extent from the roughness values $w_{FS}$, $w_{ME}$.
In the figure we have plotted $P_{ran}$ from the same
RFN configurations as $P_O$, with $A_{FS}= A_{ME}=7.5$, which is
a rather large value to be realistic, hence $P_O \not\equiv
P_{ran}$. In the figure $P_{ran}$ has a slope $\simeq -0.2$ while the
asymptotic scaling according to the random deposition argument
should be $P_{ran} \sim L^{-1/3}$ . For small systems the average
overlap is very large, of the order of 0.5 in the particular case
studied here.

Fig.~\ref{fig7}(b) shows the average size $\langle s
\rangle$ of overlapping clusters. The overlapping cluster is defined
as the number of the neighboring common $(x,z)$ coordinate pairs.  The
overlap cluster size saturates at $\langle s \rangle \simeq 8.5$.
Fig.~\ref{fig7}(c) shows the average total length of the overlap in
configurations, which do have an overlap, i.e.  $(\sum s)/N_O$, where
$N_O = P_ON$. One may write $(\sum s)/N_O = (\langle s \rangle N_s
N_O)/N_O = \langle s \rangle N_s$, where $N_s$ is the number of
overlap clusters in a system, which has overlaps. By assuming that
$N_s \sim L$, and $\langle s \rangle =$const, we get $(\sum s)/N_O
\sim L$, which is demonstrated in the figure.  Since in the large $L$
limit $(\sum s)/N_O = C_1L+C_2$, $C_2 \simeq 23$, there is a crossover
in the systems of size $L \simeq 27$, because in the small system size
limit $\sum s$ must be smaller than $L$. On the other hand $C_1 =0.15$
tells us that approximately 15\% of the length of the fracture and
yield surfaces are overlapping with each other. The scenario is that
if the fracture happens to start from the same minimum energy valley
where the DP is located, it will naturally stay localized there;
however, the associated surface stiffness is weaker and thus the
excursions. Notice the saturation of the cluster size, which agrees
with the scenario.

\begin{figure}[f]
\centerline{\epsfig{file=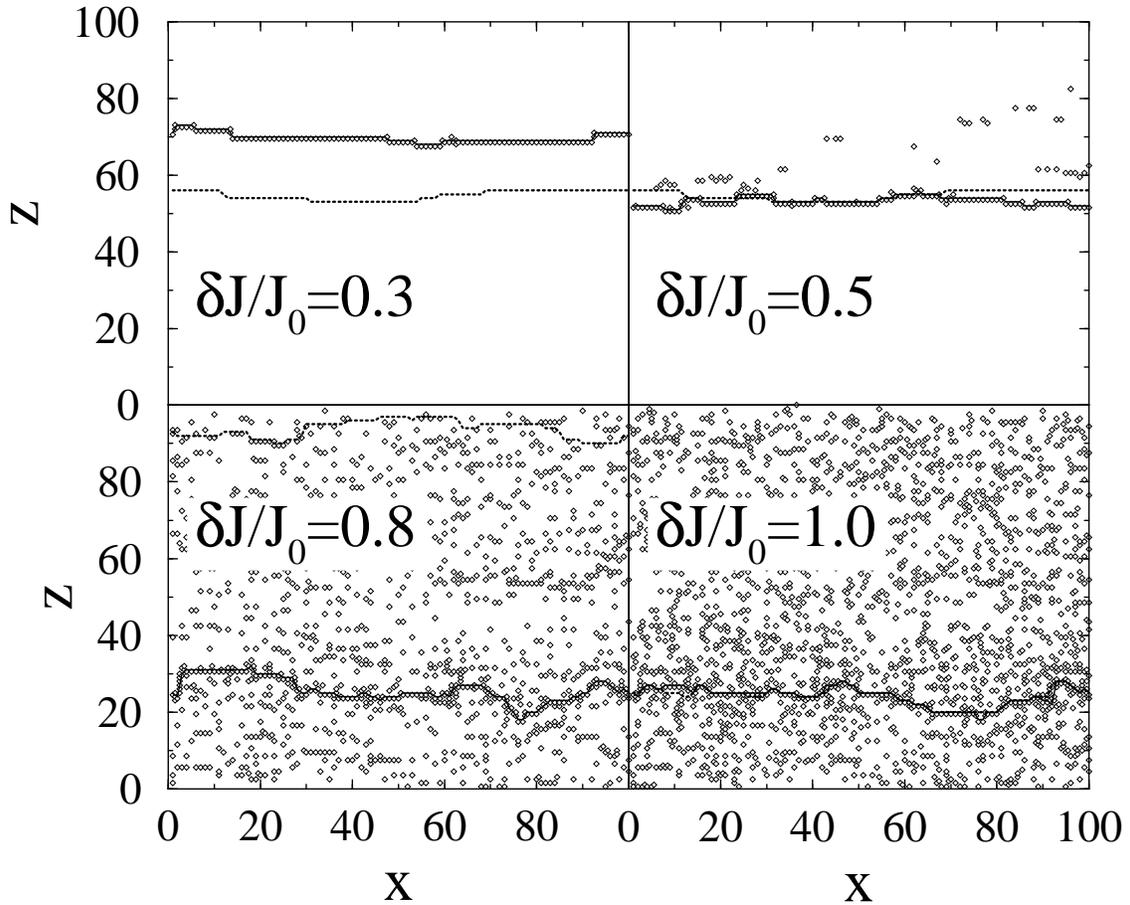,width=12cm,angle=-90}}
\caption{Examples of final damage (diamonds) and the respective
brittle failure (solid line) and yield (dotted line) surfaces for
$\delta J/J_0 =$ 0.3, 0.5, 0.8, and 1.0. The system sizes are $L^2 =
100^2$, and the random initial configuration in each system is the
same, but the $J_c$'s are rescaled with the corresponding $\delta J$.
For $\delta J/J_0=1$ the total overlap of fracture and yield surfaces
$\sum s =85$.}
\label{fignew}
\end{figure}

Fig.~\ref{fignew} shows four examples of what happens with varying
disorder strength $\delta J/J_0$ for $L=100$. The subplots demonstrate
several effects. For the weakest disorder, both the interfaces are
nevertheless ``rough'' (i.e. not flat), which shows that in spite of
a single, growing crack even the brittle fracture case can produce
a crack which fluctuates in the transverse direction. The qualitative
behavior is the same for both the cases, note that for the yield
surfaces one expects a Larkin-lengthscale on which the interfaces
look flat due to the competition between disorder and elasticity. 
With increasing disorder the
crack is finally localized in the lower part of the system - the 
threshold field is rescaled in all the cases
with the ``initial'' random number being kept constant. Meanwhile
the damage for brittle fracture grows strongly. Notice how the
yield (minimum energy) surface moves in the system as $\delta J/J_0$
is being changed. For the two cases with weakest disorder the surface
stays the same. In two of the subplots there is considerable overlap
between the fracture and the yield surfaces: $\delta J/J_0 = 1$ leads
to a total overlap of $\sum s =$ 82. 

\begin{figure}[f]
\centerline{\epsfig{file=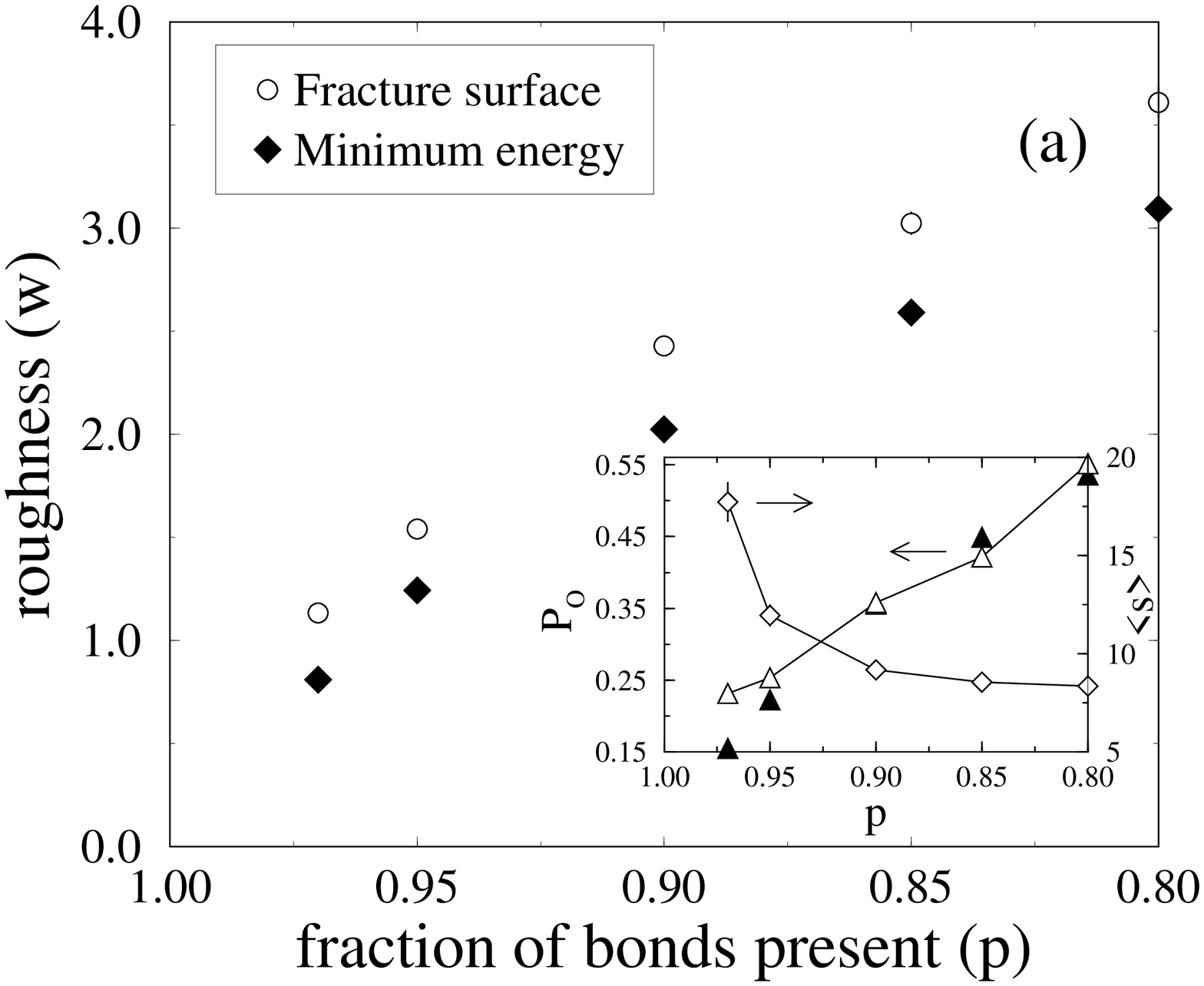,width=7cm,angle=-90}}
\centerline{\epsfig{file=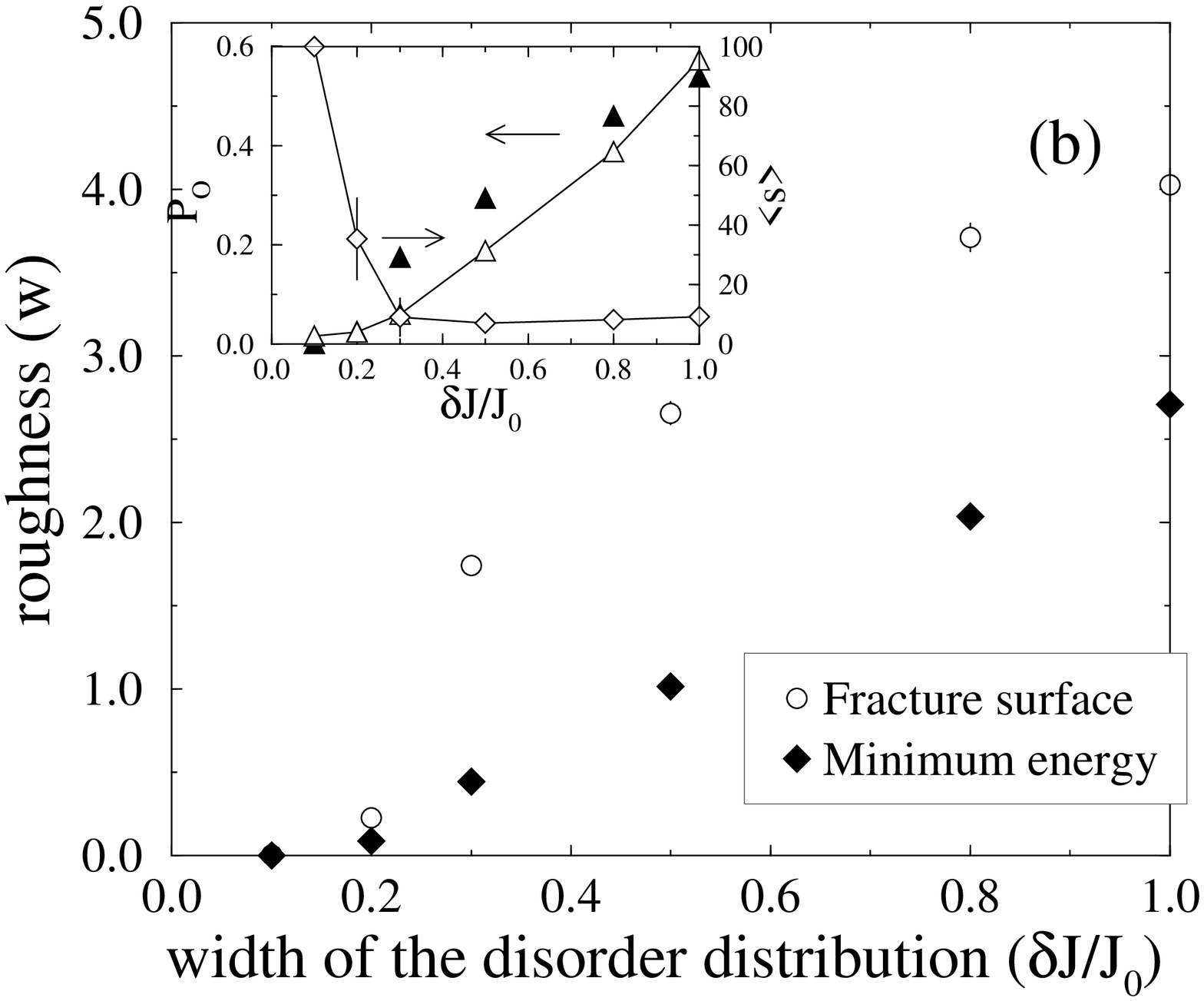,width=7cm,angle=-90}}
\caption{Interface width $w$ of the fracture and minimum energy
surfaces with varying the disorder strength for the dilution type of
disorder (a) and uniform distribution of fuse thresholds (b). The
system size $L^2=100^2$ for each system and the number of realizations
is shown in Table~\ref{tabN}. The insets show of the same systems the
fraction of the overlapping disorder configurations $P_O$, open
triangulars, $P_{ran}$ with $A_{FS}=A_{ME}=8$ in both cases, closed
triangulars, and the average overlapping cluster sizes $\langle s
\rangle$, diamonds, as a function of the disorder strength.}
\label{fig8}
\end{figure}

Fig.~\ref{fig8} shows the dependencies of the roughness and overlap
quantities on the disorder strength. Both for the dilution case,
Fig.~\ref{fig8}(a), and for the systems with the randomness from the
uniform distribution, Fig.~\ref{fig8}(b), the fracture and yield
surfaces are always rough, except for the finite size effects in the
small $\delta J/J_0$ limit. Even in this case the strong disorder
fixed point is attractive and we simply have that the system size is
smaller than the Larkin length above which the asymptotic behavior is
seen.  The roughness increases with decreasing $p$ until the
bond-percolation limit $p_c =1/2$ is reached and the surfaces become
fractals, with the corresponding hull exponent. In the insets the
average overlap $P_O$ and the average overlapping cluster size
$\langle s \rangle$ are shown.  $P_O$ increases for both type of
disorder with increasing disorder strength, except in the
infinitesimal disorder $p=1-\epsilon$ and $\delta J/J_0 = \epsilon$
limits, where it naturally diverges, the same is true for $\langle s
\rangle$ even with finite disorder. In order to compare $P_O$ with the
random deposition argument, $P_{ran}$ is plotted from the data of the
same configurations with $A_{FS}= A_{ME}=8$ showing again $P_O
\not\equiv P_{ran}$. $\langle s \rangle$ seems to saturate with
increasing disorder strength for both types of disorder around
$\langle s \rangle
\simeq 8-10$, which might be a coincidence, since one could guess it
to be disorder-type dependent.

\section{Discussion}\label{discussion}

This paper has explored the connections between brittle fracture and
minimum energy surfaces. We have given numerical support for the idea
of these being in the same universality class in two spatial
dimensions.  This argument is based on the scaling of interface width,
local roughness and the statistics of ensembles of interfaces. For
both scalar brittle fracture and perfect plasticity, or minimum
energy, interfaces these turn out to have similar scaling properties,
indicating that the brittle fracture interfaces have a roughness
exponent of 2/3 (the directed polymer one) and are also truly
self-affine. In spite of the fact that we have studied only two
types of disorder distributions, we nevertheless believe that
the numerics points to a picture of asymptotically rough cracks
in spite of the stress-enhancement effects, that one would expect
to play a role in brittle fracture with a large, dominating crack.
Such is the case in particular for the dilution-type, relatively
weak disorder, a main part of our study. Notice that for weak
disorder even minimum energy surfaces tend to be relatively flat
as the amplitude of the roughness is small.

\begin{figure}[f]
\centerline{\epsfig{file=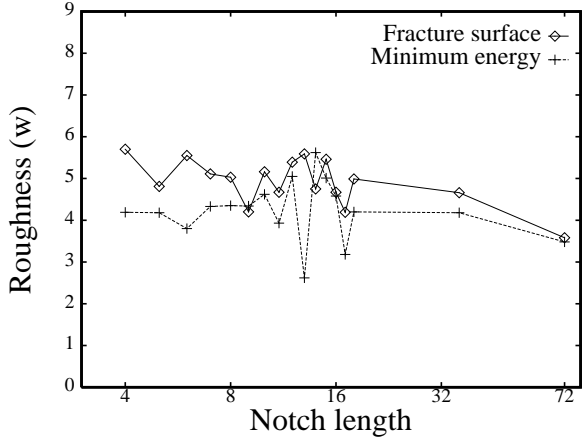,width=6cm,angle=-90}}
\caption{Roughness of fracture and minimum energy surfaces in a
notched sample with varying the notch length. The system sizes
$L^2=100^2$ and the disorder is dilution type with $p=0.8$.}
\label{fig9}
\end{figure}

The results presented earlier were obtained for systems that were
governed by extreme scaling-type arguments. The role of the disorder
can be tested in another way, more relevant to standard fracture
surface experiments, by introducing a notch, or a row of pre-failed
fuses.  The current distribution around the crack tip is a combination
of the enhancement effect of the crack plus the additional
fluctuations created by possible off-path or fracture process zone
damage. Born-model simulations by Caldarelli et al.~\cite{Calda} show
that self-affine cracks can avoid surface tension effects, i.e. they
do not straighten out, if started from point seeds. In our case the
question becomes if the effective surface tension of the notch plus
the grown crack wins over disorder, remembering that the stress
enhancement for a symmetric crack is largest on-axis. This is a
necessary mechanism for any self-affine behavior, whether of the
minimum energy surface universality class or not.  Fig.~\ref{fig9}
shows the interface roughness of yield and brittle fracture surfaces
for a fixed system size and varying notch length. It transpires that
for both ME and brittle fracture surfaces the notch effect does not
equal to flattening. Note the earlier arguments that yield surfaces
have a {\it higher} surface tension. This is again due to the memory
effect, which renormalizes the thresholds in the tangent problems so
that they are the smallest on the crack axis.

There are several experimental indications of the conclusion that
brittle fracture interfaces exhibit self-affine behavior, with a
roughness exponent close if not equal to the perfect plasticity
one. Experiments done on real materials can bridge the gap between the
two extreme limits. For such studies the expected behavior would in
2D, assuming slow failure, be self-affine as well. The extraction of
the roughness exponent has here been done using the local width as a
measure. For ensembles of experiments one should note the statistical
implications of the scaling of the roughness distribution width and of
the shape of the width probability distribution.  The relative
``irrelevance'' of a notch hints about the possibility of pinning
center -like scaling properties as should be true for the perfect
plasticity case: the notch pins the final crack with certainty if it
is large enough.

Finally we note that there is no rigorous theoretical argument that
would explain why brittle fracture seems to follow ME-type
scaling. Indeed we have here studied only the scaling of the interface
roughness, and the aspect of interface energetics in terms of e.g. the
energy fluctuation exponent $\theta$ has been left aside. Notice that
the bare strength properties are governed by logarithmic effects in
the case of the brittle failure. For brittle fracture interfaces to result
from global optimization the initial failure thresholds have to be
renormalized by the correlations which the stress-intensities of the
crack history induces. For such extremal statistics processes no
theory exists for the time being, unlike for the case of the quenched
Laplacian breakdown model for which one can use real-space
renormalization \cite{cafi}.

\medskip
\centerline{\bf Acknowledgements}
\medskip

\noindent
The authors would like to thank the Finnish Cultural Foundation (VR)
and the Academy of Finland for financial support (the Matra program,
VR separately). 

%%%%%%%%%%%%%%%%%%%%%%%%%%%%%

%%%%%%%%%%%%%
%FIGURES
%%%%%%%%%%%%%

\end{multicols}
\widetext

\begin{multicols}{2}[]
\narrowtext

\begin{table}
\caption{The number of realizations $N$ performed in simulations 
for exactly the same randomness of brittle failure and plasticity.}
\label{tabN}
\begin{tabular}{l c c c c}
 & \multicolumn{2}{c}{dilution, $p$}&\multicolumn{2}{c}{uniform, $\delta J/J_0$}\\
$L$ & $0.8$ & $0.85-0.97$ & $1$  & $0.1-0.8$  \\
\tableline
$10, 20$     & 760 &     &     &     \\
$30-90$      & 760 &     & 66  &     \\
$100$        & 370 & 537 & 248 & 250 \\
$200$        & 370 &     &     &     \\
$275, 350$   & 250 &     &     &     \\
\end{tabular}
\end{table}

\end{multicols}
\widetext

\end{document}